# The Demonstration Model of the ATHENA X-IFU Cryogenic AntiCoincidence Detector

M. D'Andrea[1] • C. Macculi[1] • G. Torrioli[2] • A. Argan[1] •
D. Brienza[1] • S. Lotti[1] • G. Minervini[1] • L. Piro[1] • M. Biasotti[3] •
L. Ferrari Barusso[3] • F. Gatti[3] • M. Rigano[3] • A. Volpe[4] •
E. S. Battistelli[5,6,1]

[1]INAF/IAPS, Via del Fosso del Cavaliere 100, I-00133 Roma, Italy
[2]CNR/IFN Roma, Via Cineto Romano 42, I-00156 Roma, Italy
[3]University of Genoa, Via Dodecaneso 33, I-16146 Genova, Italy
[4]ASI, Via del Politecnico, I-00133 Roma, Italy
[5]Sapienza University of Rome, Piazzale Aldo Moro 5, I-00185, Rome, Italy
[6]INFN - Sezione di Roma, Piazzale Aldo Moro 5, I-00185, Rome, Italy

**Abstract** The Cryogenic AntiCoincidence Detector (CryoAC) of ATHENA X-IFU is designed to reduce the particle background of the instrument and to enable the mission science goals. It is a 4 pixel silicon microcalorimeter sensed by an Ir/Au TES network. We have developed the CryoAC Demonstration Model, a prototype aimed to probe the critical technologies of the detector, i.e. the suspended absorber with an active area of 1 $cm^2$; the low energy threshold of 20 keV; and the operation connected to a 50 mK thermal bath with a power dissipation less than 40 nW. Here we report the test performed on the first CryoAC DM sample (namely the AC-S10 prototype), showing that it is fully compliant with its requirements.

M. D'Andrea • C. Macculi • G. Torrioli et al.

## 1 Introduction

ATHENA [1] is a large X-ray observatory, selected by ESA for launch in 2031. One instrument of the payload is the X-IFU [2], a cryogenic spectrometer providing spatially resolved high-resolution spectroscopy. The core of the instrument is a 4 kilo-pixel TES array operated at a 50 mK thermal bath temperature. Since the particle background expected in the spacecraft orbit would strongly limit the instrument sensitivity, advanced reduction techniques have been adopted to reduce it by a factor ∼50 and reach the scientific requirement of 0.005 cts/cm$^2$/s/keV in the 2-10 keV energy band (see [3] for details). This is mandatory to allow the observation of faint extended sources and enable many core science objectives of the mission, e.g. the study of the hot plasma in galaxy cluster outskirts [4]. Most of the background reduction is achieved thanks to the Cryogenic AntiCoincidence detector (CryoAC), a 4 pixel TES microcalorimeter placed less than 1 mm below the TES array. The CryoAC is an instrument-inside-the-instrument, with independent electronics and a dedicated data processing chain. It shall have a wide energy band (from 6 keV to 1 MeV - To Be Confirmed - ) and a low deadtime (1%), while respecting several constraints to ensure mechanical, thermal and electromagnetic compatibility with the TES array.

The X-IFU development plan foresees to build an instrument Demonstration Model (DM) before the mission adoption, in order to qualify the instrument critical technologies. In this respect, we have developed the CryoAC DM, which is aimed to probe the following requirements: suspended absorber with an active area of 1 cm$^2$; low energy threshold less than 20 keV; operation connected to a 50 mK thermal bath; power dissipation at 50 mK less than 40 nW.

## 2 The CryoAC Demonstration Model

The CryoAC DM (namely AC-S10 sample, Fig. 1) is a single pixel detector based on a large area (1 cm$^2$) Silicon absorber (see [5] for fabrication details). To obtain a well-defined conductance towards the thermal bath, the absorber is connected to a gold-plated silicon rim through 4 narrow silicon beams (100 x 1000 μm$^2$), achieving a suspended structure. This is sensed by a network of 96 Ir/Au TES connected in parallel, and read out by a VTT K4 SQUID [6] operating in the standard Flux Locked Loop (FLL) configuration. The TES network is designed to achieve efficient athermal phonons collection, and it features anti-inductive Nb wirings to limit the electromagnetic coupling with the TES array (see [5] for details).

# The Demonstration Model of the ATHENA X-IFU CryoAC

Platinum heaters are also embedded on the absorber for calibration and diagnostic purposes. The heaters are placed in proximity of the silicon beams, and they are connected in parallel by niobium wiring, with a resulting equivalent resistance Rheater = 307 Ω. If necessary, they could be also used during operations to increase the local temperature and reduce the current needed to operate the TES network, limiting crosstalk effects on the TES array. The main parameters of the detector are summarized in Table 1.

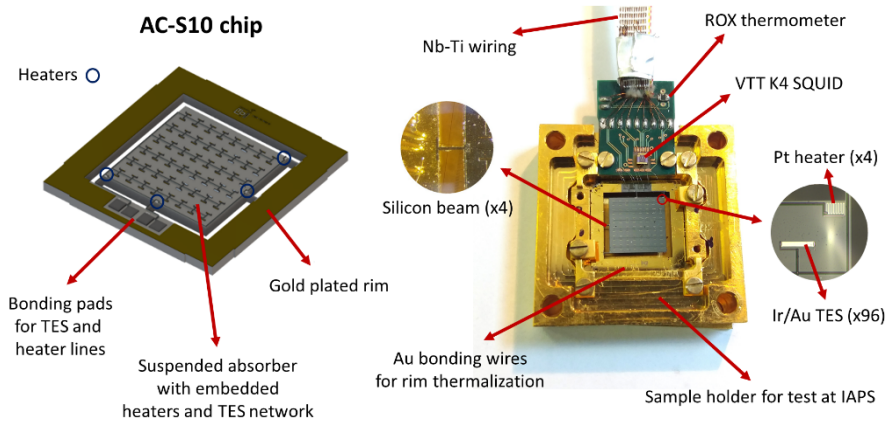

**Fig. 1** The CryoAC DM (AC-S10 sample). *Left:* Sketch of the detector chip. The blue circles indicate the position of the onboard heaters. *Right:* The detector assembly. Note that the chip gold plated rim is thermally anchored to the sample holder via gold bonding wires (Color figure online).

**Table 1** Main parameters of the CryoAC DM (AC-S10 sample)

| Parameter | Value |
| --- | --- |
| Silicon chip thickness | 525 μm |
| Total chip area | 16.6 x 16.6 mm$^2$ |
| Suspended absorber area | 10.0 x 10.0 mm$^2$ |
| Beams dimensions | 1000 x 100 μm$^2$ |
| Ir/Au TES size (x 96) | 50 x 500 μm$^2$ |
| Ir/Au TES thickness | 320 nm (Ir 240 nm + Au 80 nm) |
| Pt heater resistance | 307 Ω |
| Pt heater thickness | 50 nm |

**3 Thermoelectric characterization**

The sample has shown a narrow superconductive transition (Fig. 2 inset), with a critical temperature $T_C$ = 106 mK and a normal resistance $R_N$ = 31.3 mΩ. To evaluate the thermal conductance of the system we have used the Pt on-board heater, measuring the thermal power needed to drive the unbiased TES inside the transition (T = $T_C$) as a function of the thermal bath temperature (Fig. 2). The data have been fitted with the standard power-law model describing the power flow to the heat bath [7]:

$$P_H = k(T_C^n - T^n) \qquad (1)$$

where *k* and *n* are constants determined by the nature of the thermal link. The best-fit parameters are reported in the plot. Note that we have found $n \sim 4$, as expected for a phonon-mediated material [7], indicating that we have actually measured the power flowing through the silicon beams.

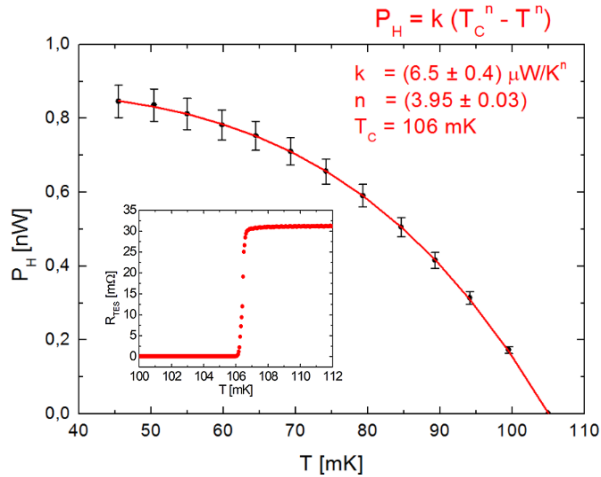

**Fig. 2** Power injected on the absorber to bring the unbiased TES into the transition as a function of the thermal bath temperature. *Inset* The TES superconducting transition (Color figure online).

The final measured thermal conductance is:

$$G = k\, n\, T^{(n-1)} = 34\, nW/K \; at \; T = T_C = 106\, mK \qquad (2)$$

## The Demonstration Model of the ATHENA X-IFU CryoAC

The characteristic I-V curves of the sample at different thermal bath temperatures are reported in Fig. 3 - Left. The current flowing through the TES ($I_{TES}$) is measured as a function of the current injected to the bias circuit ($I_{BIAS}$), where the TES is connected in parallel with a $R_S = 0.5$ mΩ shunt resistor. The slope of the curves in the TES superconducting region is affected by a current-dependent parasitic resistance ($R_P \sim 20$ mΩ at maximum) appearing in series with the TES for $I_{TES} > \sim 10$ μA. We are investigating its nature, but we report that it does not prevent to operate the detector in compliance with its requirements.

Thanks to the on-board heater, we have been able to acquire another family of characteristic curves, injecting different amounts of power on the absorber while keeping fixed the bath temperature at $T_B = 50$ mK (Fig. 3 - Right). These curves define a large set of points in which it is possible to operate the detector with the bath at 50 mK. Note that the use of the heater allows a reduction in the current needed to bias the TES inside the transition (the more the power injected by the heater, the less the current needed to bias the TES). This has the benefit of reducing also the current flowing through the shunt resistor, and consequently its power dissipation, which is a dominant contribution in the total power dissipated on the cold stage. The cost for this is to reduce the loop gain of the system, slowing down the detector.

In Table 2 are reported as example the specification of three analogous working point of the detector ($R_{TES} = 10\% R_N \sim 3$ mΩ), obtained at different heater powers. Note that the more the power injected by the heater, the less the total power dissipation at cold and the less the loop gain.

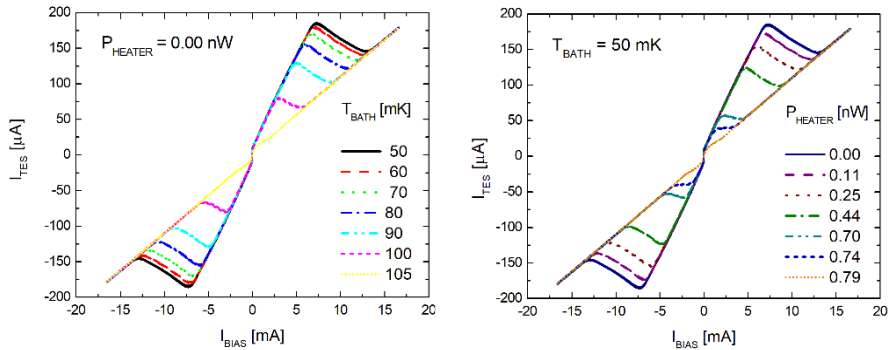

**Fig. 3** Characteristic I-V curves of the sample. *Left* "Standard I-V": $I_{TES}$ vs $I_{BIAS}$ sweeping $T_B$. *Right* "Operative I-V": $I_{TES}$ vs $I_{BIAS}$ at $T_B = 50$ mK sweeping the power injected by the heater. (Color figure online)

**Table 2** Specifications of three possible working point of the detector at $T_B$=50 mK. In each point $R_{TES}$ = 10% $R_N$ ~ 3 mΩ. $P_H$ is the power injected by the heater. $P_{SH}$, $P_{TES + PAR}$ and $P_{SQUID}$ are the power dissipated by the shunt resistor, by the TES and the parasitic resistance, and by the SQUID. $P_{TOT} = P_H + P_{SH} + P_{TES + PAR} + P_{SQUID}$ is the total power dissipation at 50 mK. The third line (in bold) corresponds to the working point used for the measurements reported in Sect. 4 and 5.

| # | $P_H$ [nW] | $I_{BIAS}$ [mA] | $I_{TES}$ [µA] | $P_{SH}$ [nW] | $P_{TES + PAR}$ [nW] | $P_{SQUID}$ [nW] | $P_{TOT}$ [nW] | Loop Gain |
|---|---|---|---|---|---|---|---|---|
| 1 | 0.00 | 8.10 | 181 | 30.1 | 0.76 | 1.5 | 32.3 | 195 |
| 2 | 0.25 | 6.75 | 150 | 20.7 | 0.51 | 1.5 | 23.0 | 135 |
| **3** | **0.74** | **0.80** | **39** | **0.29** | **0.02** | **1.5** | **2.6** | **9** |

## 4 Operation at 50 mK thermal bath and low energy threshold

The detector has been operated in the third working point in Table 2, with a total power dissipation $P_{TOT}$ = 2.6 nW at 50 mK, fully compliant with the DM requirement ($P_{TOT}$ < 40 nW).

To probe the low energy threshold the sample has been illuminated by a $^{55}$Fe source (6 keV photons), at a count rate of ~ 10 cts/s. Simultaneously, the sample has been also stimulated injecting fast (1 µs) square current pulses into the on-board heater ($I_{H,PULSE}$ = 4.3 µA), generating thermal pulses with an energy E = 35 keV at the rate of 2 cts/s (Fig. 4 Left).

The acquired pulses show a rise time $\tau_R$ ~ 30 µs and a thermal decay time $\tau_D$ ~ 5 ms. The energy spectrum is shown in Fig. 4 - Right. It has been obtained from the Pulse Height spectrum, calibrating the energy scale to the 6 keV line. Note that the heater pulses line is properly centered around the energy of 35 keV, enabling the use of the heater as calibration pulse generator and probing the linearity of the detector response in this energy range. The measured low energy threshold is $E_{threshold}$ = 2.75 keV, fully compatible with the DM requirement ($E_{threshold}$ < 20 keV). It corresponds to the 5 sigma level over the noise, which has been evaluated from the baselines of the acquired pulses. The sample shows also some spectroscopic capability (ΔE = 1.30 keV @6 keV). Although the CryoAC is not aimed to perform spectroscopy (and there is not a related requirement on the DM), this could provide some scientific return for the X-IFU (see [8] for details).

**The Demonstration Model of the ATHENA X-IFU CryoAC**

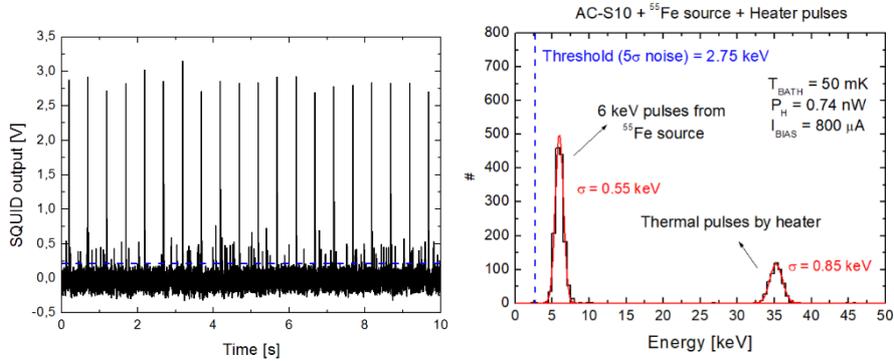

**Fig. 4** *Left* Functional test of AC-S10 operated at $T_B$=50 mK. The blue dashed line is the low energy threshold level (5 sigma noise). *Right A*cquired energy spectrum fitted by gaussian functions (red line). (Color figure online)

**5 Pulse response up to the saturation regime**

We have finally used the on-board heater to stimulate the detector in its whole band, in order to test the response up to the saturation regime. Also for this measurement the detector has been operated in the third working point in Table 2. The acquired pulses are shown in Fig. 5 - Top. For each pulse we have measured the pulse height PH (Fig. 5 - Bottom Left) and the recovery time $t_{REC}$ (Fig. 5 - Bottom Right). The recovery time is here defined as the sum of the pulse decay time (i.e. the characteristic time at which the signal is reduced to 1/e times its maximum value) and the time in which the pulse is saturated (see Fig. 5 - Bottom Right, inset).

The study of the detector behavior in saturation is important since it is related to its Dead Time, which is one of the main requirement driving the CryoAC design (DT<1%). From the plots in Fig. 5 we can evaluate for the CryoAC DM a detector saturation threshold of ~ 1.3 MeV, and assess that the recovery time in saturation grows roughly logarithmic with the deposited energy. Although the DM does not have requirements related to these aspects, these kind of studies are useful in view of the CryoAC Flight Model design.

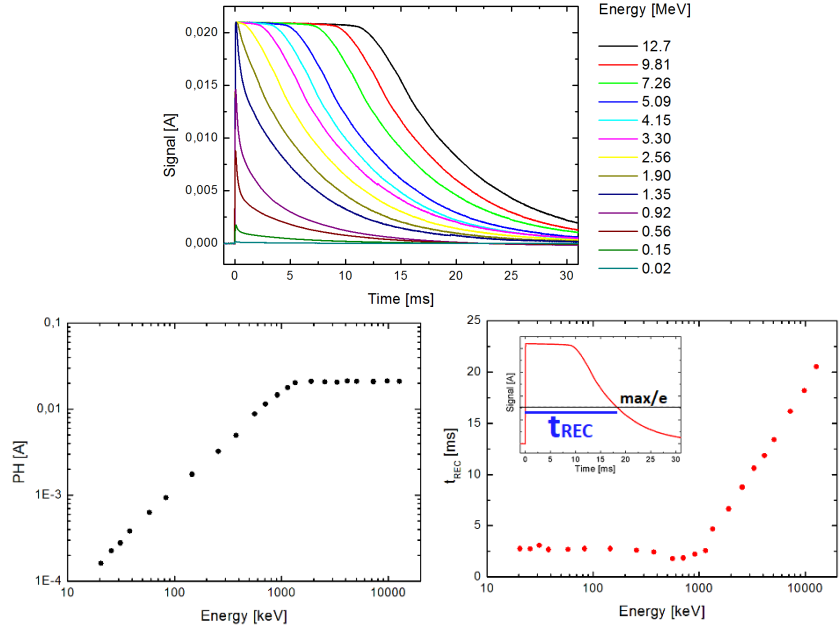

**Fig. 5** *Top* Pulses acquired stimulating the detector by the on-board heater. *Bottom Left* Pulse Height as a function of the energy. *Bottom Right* Recovery time as a function of the energy. Inset: Evaluation of the recovery time for saturated pulses. (Color figure online)

## 6 Conclusions

We have reported the main results of the preliminary test activity performed on the CryoAC DM (AC-S10 sample). The detector is compliant with its requirements, showing a total power dissipation of 2.6 nW at 50 mK (DM requirememt: 40 nW) and a low energy threshold < 3 keV (DM requirement: 20 keV). The on-board heater turned out to be useful to optimize the detector operation, reducing the total power dissipation at cold. Beside this, it has allowed us to properly measure the thermal conductance of the system and test the detector response up to the saturation regime.

After this first stand-alone test performed at INAF/IAPS, the CryoAC DM has been delivered to SRON, for integration at the chipset level with the TES array. This will be the first compatibility test for the two detectors, representing a milestone on the path towards the X-IFU development.

# The Demonstration Model of the ATHENA X-IFU CryoAC

**Acknowledgements** This work has been supported by ASI contract n. 2018-11-HH.0 and ESA CTP contracts n. 4000116655/16/NL/BW and n. 4000114932/15/NL/BW. The authors would like to acknowledge the SRON X-IFU FPA team for the support in the development of the CryoAC DM setup, and Giorgio Amico (Sapienza University of Rome) for the production of CryoAC DM brackets.